\documentclass[secnumarabic,amssymb, nobibnotes, aps, prd]{revtex4}%
\usepackage{graphicx}
\usepackage{dcolumn}
\usepackage{bm}
\usepackage{amsmath}%
\usepackage{amsfonts}%
\usepackage{amssymb}

\begin{document}

\title{Evidences of high energy protons with energies beyond 0.4 GeV in the solar particle
spectrum as responsible for the cosmic rays solar diurnal anisotropy}

\author{C. E. Navia, C. R. A. Augusto, M. B. Robba and K. H. Tsui}
\address{Instituto de F\'{\i}%
sica Universidade Federal Fluminense, 24210-346,
Niter\'{o}i, RJ, Brazil} 

\date{\today}
\begin{abstract}
Analysis on the daily variations of cosmic ray muons with $E_{\mu}\geq 0.2\;GeV$ based on the data of two directional muon telescopes at sea level and with a rigidity of response to cosmic proton spectrum above 0.4 GV is presented. The analysis covers two months of observations and in 60\% of days, abrupt transitions between a low to a high muon intensity and vice-verse is observed, the period of high muon intensity is from $\sim 8.0h$ up to $\sim 19.0h$ (local time) and coincides with the period when the interplanetary magnetic field (IMF) lines overtake the Earth. This behavior strongly suggest that the high muon intensity is due to a contribution of solar protons (ions) on the muon intensity produced by the galactic cosmic rays, responsible for the low muon intensity. This implies that the solar particle spectrum extends to energies beyond 1 GeV. We show that this picture can explain the solar daily variation origin, and it is a most accurate scenario than the assumption of corotating galactic cosmic ray with the IMF lines, specially in the high rigidity region. Obtained results are consistent with the data reported in others papers. Some aspects on the sensitivity of our muon telescopes are also presented.

 \end{abstract}


\maketitle

\section{Introduction}

Galactic cosmic rays (GCR) beyond the heliosphere region are considered to be temporally and spatially isotropic at least over large timescales. They enter into the heliosphere due to random motions and they are subjected to a number of transport effects such as diffusion, adiabatic cooling, convection and drift \cite{parker63}. The relative importance of these processes varies with particle properties, such as energy, timescales and of course the position in space. The size and structure of the modulation volume is assumed to be approximately a sphere with a radius of $\sim 100$ AU.

On the other hand, the term ``solar energetic particles'' (SEP) include all particles in the heliosphere accelerated by processes related to solar activity, such as the anomalous cosmic ray (ACR), particles accelerated in corotating interaction region (CIR), as well as particles accelerated in solar flares and coronal mass ejection (CME). While, it is believed that particles, continuously expelled by the Sun, such as the solar wind plasma,  
have energies up to several GeVs in the tail of the solar energy spectrum     
 and only during transient solar events (i.e. solar flares) their energies can reach dozens of GeVs. So far, the particle acceleration mechanism by the Sun in a wide band of energies, specially the high energy region, are still poorly understood. 
The survey on modulation of cosmic ray particles by the interplanetary magnetic field and the correlation with solar activity began using ground based cosmic ray detectors in the thirties. Due to the complexity, details of the phenomena are still subjected to studies.

Interaction  of the primary cosmic rays with the atmosphere produce, among other things, a lower energy secondary nucleons, in particular, neutrons that are not slowed by ionization loss. These secondaries fall in the energy range of a few hundred MeV up to about $\sim 1$ GeV. These nucleons in turn produce further nuclear interaction, either in the atmosphere or in lead target materials surrounding the detectors, in most of cases the so called neutron monitors (NMs). The interaction rate may be measured most conveniently and reliably by detecting the reaction products in neutrons rather than by detecting the charged fragments directly.

The NMs worldwide network starting from 1954 by Simpson \cite{simpson54} has shown excellent performances because the intensities are recorder to several geomagnetic cutoffs and anisotropies and other characteristic can be better known. One of the main obtained results using NMs is the long term variation, the cosmic ray
intensity increase and decrease with the solar cycle. They are in anti-correlation with the number of solar spots. On average, every eleven years, solar activity is high and therefor cosmic rays are deflected stronger than during
minimum solar activity. Consequently, when the Sun is active, fewer galactic cosmic rays reach Earth's atmosphere.  The anti-coincidence guard counting rate of GCRs in spacecrafts (1972-2002) have confirmed this long term variation
\cite{richardson04}.

On the other hand, temporal variations of the cosmic ray intensity as an abrupt intensity decrease at ground level were observed already in the thirties by Forbush \cite{forbush37}. These ``Forbush'' events are associated to the passage at the Earth's vicinity of a disturbance (shock and plasma) causing a shielding effect. At least in the case of large Forbuch events, the disturbance ``eject'' is emitted by the Sun during the coronal mass ejection process. In addition, near the Sun's equatorial plane high and low speed solar wind flows interact. This interaction is known as corotating interaction region (CIR). There are forward and reverse shocks bounding these regions, which are known to modulate galactic cosmic rays.  Abrupt depressions in the GCR intensity were also observed in the vicinity of the maximum solar wind speed by Iucii and coworkers \cite{iucci79} analyzing neutron monitor data and later confirmed by spacecraft experiments. These cosmic ray modulations are associated with corotating high-speed streams and CIRs.    

Another important result obtained by ground based cosmic ray detectors, NMs, as well as underground muon telescopes is the short term variation, known as the solar diurnal variation or daily anisotropy of the cosmic ray intensity. It has been observed in detectors located at different global sites and in a wide range of the cosmic ray spectrum, rigidities between 1 GV to 500 GV. The solar diurnal variation is attributed to the bulk streaming of the cosmic ray gas caused by the corotating interplanetary magnetic field that is rigidly attached to the Sun and it is related in terms of diffusion, convection, and drift of GCR in the IMF \cite{forman75}. However, at high rigidities (above 10-20 GV) the standard convection-diffusion picture for GCR is inaccurate and probably inapplicable, because the concept of a diffusive streaming breaks down \cite{kota97}. Consequently, the exact nature of galactic cosmic ray contribution to the solar daily anisotropy is not yet clear. 

A completely different sort of information on primary cosmic ray comes from the Tupi experiment, located at sea level. Starting from April of 2007 this experiment consists of two identical directional muon telescopes, 
 constructed on the basis of plastic scintillators.
 One of them with vertical orientation and another one with an orientation of 45 in relationship of the vertical (zenith) and pointing at the west. Both with an effective aperture of $65.6\;cm^2\;sr$.  The rigidity of response of these detectors to cosmic proton spectrum is above 0.4 GV, allowing registration of muons with $E_{\mu}\geq 0.2\;GeV$. The daily variation of muon intensity, in most cases, consists of a high muon intensity observed between 9 hours and 18 hours (local time), and a low muon intensity, up to ten times smaller than the high muon intensity, is observed in the remaining 13 hours. This behavior on the daily variation is subjected to several fluctuations of different sort, such as magnetic disturbances in the heliosphere inside the interplanetary space and near the Earth.
 
 Due to the abrupt transition between the low to high muon intensity and vice-verse, as well as, the region of high muon intensity $\sim 8.0h-19.0h$ (local time) coincides with the period where the IMF lines overtake the Earth, we argue that the high muon intensity is due to a contribution of solar protons (ions) on the muon intensity and the galactic cosmic rays are responsible of the low muon intensity. 
If this picture is right the solar particle spectrum extend to energies beyond 1 GeV, and probably up to $\sim 500$ GV in the tail of the spectrum, because the daily variation has been observed besides in this high rigidity range.
This scenario can explain in a natural way the solar daily anisotropy, without the co-rotating  
assumption of galactic cosmic rays with the IMF inside the standard convection-diffusion picture, hard to be applied in the high rigidity region.
 
Our conclusion is on the basis of the average daily muon intensity with $E_{\mu}\geq 0.2\;GeV$ produced in the Earth's atmosphere by protons (ions) with rigidity above 0.4 GV over a period of two months, and coincides with the actual period of minimum solar activity. Obtained results are in agreement with the data reported in others papers.

\section{The Tupi experiment}

The Tupi experiment is a muon tracking telescope located at sea level, and whose coordinates are: $S22^054'33''$ latitude and $W43^008'39''$ longitude. Observations on the muon flux  and muon enhancements in association with solar transient events have been reported \cite{augusto03, navia05, augusto05}.
Starting from April of 2007 we have initiated the Phase II of the Tupi experiment with
two identical muon telescopes on the basis of plastic scintillators, one of them with vertical orientation and another one with an orientation of 45 in relationship of the vertical (zenith) and pointing at the west, both with an effective aperture of $65.6\;cm^2\;sr$. The telescopes are inside a building under two flagstones of concrete and allowing registration of muons with $E_{\mu}\geq 0.2\;GeV$ required to penetrate the two flagstones, Fig.1 summarized the situation.
The rigidity of response of these detectors to cosmic proton (ions) spectrum is given by the local geomagnetic cutoff 0.4 GV. This low value is due to the Brazilian magnetic anomaly (see section 4).

The directionality of the muon telescopes is guaranteed by a veto or anti-coincidence guard as is shown in Fig.2.
The vertical telescope has a veto or anti-coincidence guard, using a detector of the inclined
telescope and vice-verse. Therefore, only muons with trajectories close to the telescope axis are registered. The data acquisition is made on the basis of the Advantech PCI-1711/73 card
with an analogical to digital conversion at a rate of up to 100 kHz.

\section{The muon component at sea level}

The primary cosmic ray particles (i.e. protons and nuclei) can be inferred through the detection of muons 
by telescopes at ground and underground levels.  The upper layers of the Earth's atmosphere is bombarded by a flux of cosmic primary particles. The chemical composition of this primary cosmic particles depend on the energy region. In the low energy region (above 1 GeV to several TeV) the dominant particles  are protons ($\sim 80\%$). The primary cosmic rays collide with  the nuclei of air molecules and produce an air shower of particles that include nucleons, charged and neutral pions, kaons
etc. These secondary particles then undergo electromagnetic and nuclear interactions to produce yet additional particles
in a cascade process. Of particular interest is the fate of charged pions, $\pi^{\pm}$, produced in the cascade. Some of these will interact via the strong interaction with air molecule nuclei  but others will spontaneously decay via the weak interaction into a muon, $\mu^{\pm}$, plus a neutrino or anti-neutrino, $\nu_{\mu}$, following the scheme
\begin{equation}
\pi^{\pm}\rightarrow \mu^{\pm} \nu_{\mu}.
\end{equation}
Muons are quite penetrating and they can reach the ground, enter the laboratory through the walls or roof of the building, and be detected with a suitable apparatus and very high energy muons reaches underground levels.   

The expected energy spectra of muons \cite{green79,lipari93} at sea level at two zenith angles, $\theta=0^0$ and $\theta=45^0$ are shown in Fig.3. For muons with energies around $E_{\mu}\sim 2\;GeV$ the vertical intensity $\theta=0^0$ is twice of the intensity at $\theta =45^0$. This behavior is a consequence of the zenith angle distribution of muons in the GeV region, to be related as $\cos^2 \theta$ with the value of $0.5$ for $\theta=45^0$.

Exactly, this is the reason for the Tupi experiment using two identical telescopes, a vertical one and another one inclined in an angle of $45^0$. We have observed that in average the counting rate in the vertical telescope is twice the counting rate  in the inclined telescope as is shows in Fig.4 (upper panel), where the 5 min muon counting rate, observed in both telescopes vertical and inclined, before and after pressure correction and for two consecutive days 3th and 4th April 2007, is presented. 
On the average, the counting rate on the vertical direction is two times higher than the counting rate on the inclines ($45^0$) direction, as the two telescopes have the same aperture $65.6\;g/cm^2$, the intensity in the vertical direction is twice than intensity in the inclined direction.
This behavior is used as a ``quality control'' in the output time series. 
From this figure, it is possible to see that
the influence of the atmospheric pressure variation (see Fig.4 lower panel) on the muon intensity at sea level is not relevant.

\section{The Tupi muon telescopes sensitivity}

Ground-level solar cosmic ray events are usually observed by high latitude neutron monitors at relatively low rigidities ($\geq 1$ GV) and in most cases the ground-level events are linked with solar flares of high intensity
whose prompt X-ray emission is cataloged as X-class (above $10^4Wm^{-2}$) Evidently, the solar flare detection at ground
depends on several aspects such as a good magnetic connection between the Sun and Earth.
In the new array (Phase II) of the Tupi experiment,
constituted by two directional muon telescopes and during the present survey for diurnal anisotropies, it were found  several muon enhancements. The 
event presented here is clearly distinguishable above the muon background, 
with statistically significance of $17\sigma$ in temporal coincidence with a flare of small scale whose prompt X-ray emission is cataloged by the GOES11 as C3 class on April 16, 2007 at 22.8 h UT (see Fig.5). 
This is the first detection of a ground-level solar cosmic ray event linked with the impulsive or prompt emission of energetic particles emitted during a very small solar flare (it is the record in the detection at ground of the smallest solar flare). The Tupi telescope had already detected before events tied with the gradual emission via CME of a  flare of small scale, see ref. \cite{navia05,augusto05}. 

The high sensitivity of the Tupi telescopes can be in part a consequence of the Brazilian magnetic anomaly, with a minimum at 26S, 53W, which is very close to the position 22S, 43W where the Tupi experiment is located. The lowest magnetic field in the region of the anomaly at ground is three times lower than the magnetic field at polar regions.  Then, the area of the magnetic anomaly is like a funnel for incoming charged particles from space. Fig.6 summarized the situation, where the contours of the omni-directional proton intensity ($E_p>10$ MeV) around the anomaly region is plotted \cite{nichitiu}. 

The Brazilian magnetic anomaly is responsible for the low rigidity response ($\sim 0.4$ GV) of the Tupi telescopes to incoming protons (ions) from space. Without the magnetic anomaly, the geomagnetic cutoff at latitude where the Tupi telescopes are located is around $\sim 10\;GV$. We believe that the presence of the Brazilian magnetic anomaly, on the place of the Tupi experiment is responsible at least partly for the results obtained about the daily variation on the muon intensity and that will be presented in the section 6. A more accurate conclusion on this subject require further study.

\section{The solar diurnal anisotropy}

The solar daily variations known also as the diurnal solar anisotropy of the cosmic ray intensity have been observed by ground and underground based detectors, covering a wide range of the cosmic ray spectrum, rigidities between 1 GV to 400 GV. The anisotropy reflects the local interplanetary cosmic ray distribution. It is widely believed that the anisotropy arises when the galactic cosmic ray particles (GCR) co-rotate with solar wind stream following the Interplanetary Magnetic Field (IMF) lines, and it is related in terms of diffusion, convection and drift of galactic cosmic ray in the IMF.

According to Forman and Gleeson  \cite{forman75} at 1 AU the co-rotating stream (solar wind) has a speed of order 450 $km\;s^{-1}$ (in average) and at $\sim 18h$ (local time) approximately in the same direction as the Earth's orbital motion (of 30 $km\;s^{-1}$). In other words, the co-rotating stream will overtake the Earth ``almost in the vertical'' from the direction of $\sim 18\;h$, and it is known as the phase of the anisotropy. This phase have shifted toward earlier hours $\sim 15h$ during the lower solar cycle due to drift process. So far, the drift process is still the most likely and accepted explanation of the phase shift.  

However, the solar daily variation specially observed in the high rigidity region have shown remarkable changes in phase and amplitude during long period of observation.  A review of characteristics of the observed cosmic ray diurnal variation over three decades has been reported by
Ahluwalia and Fikani \cite{Ahluwalia97}. They conclude that the maxima and minima in the amplitude and phases exhibit features related to the solar activity, and the amplitude decrease systematically with increasing value of rigidity. We have extracted from this survey the maxima and minima in the phases obtained with the Deep River ($R=16GV$) and Huancayo ($R=33GV$) Neutron Monitors (NMs) and the Embudo ($R=134GV$)  underground muon telescope (MT) for the 1965 to 1994 period, as well as, we have included the results of the Nagoya ($R=60GV$) muon telescope (MT) reporter in the Mori's review paper \cite{mori96}, for the 1971 to 1994 period.  

From these results we conclude that the low rigidity diffusive-convective stream picture predictions as $\sim18h$ and $\sim 15h$ for the maximum and minimum  phase of the anisotropy becomes inaccurate in the high rigidity region. The situation becomes critical analyzing the data of the Nagoya muon telescope since both values, minimum and maximum of the phase are completely out of two theoretical prediction. These results strongly constrain the assumption of corotating galactic cosmic ray as the origin of the anisotropy. Or at least the exact nature of galactic cosmic ray contribution to the solar daily anisotropy is not yet clear. 

The above  results show that the phase  of the solar daily anisotropy is distributed in a wide range from $\sim 9h\;LT$ up to $\sim 18h\;LT$.
This behavior of the phase distribution agree with our results on the basis of muon intensity registered at sea level by two directional muon telescopes and they are presented in the next section.

\section{Results}

Starting from April of 2007 we have initiated the Phase II of the Tupi experiment, with a survey on the daily variation of the muon intensity at sea level using two identical muon telescopes at sea level as described in section 2. The method applied here to study the cosmic ray anisotropy is based on the idea that a fixed detector scans the sky due to the Earth's rotation.

In each 0ne of these figures, Fig.7, Fig.8, Fig.9 and Fig.10, we show representative data on the daily variation of the muon intensity, observed in five consecutive days. For the particular period shown in Fig.7, we can see in the first three days on 2, 3, and 4 of May 2007, there are two levels in the muon intensity: a high muon intensity level between the $\sim 12h$ UT and $\sim 21h$ UT and correspond to 9h LT and 18h LT  and a low muon intensity in the remaining hours. Hereafter, the period of high intensity
12h to 18h will be called with the jargon ``solar window to muons''

The blue and red lines marked as Tupi 4 and Tupi 2 respectively are the muon intensities registered in the vertical direction, before and after atmospheric pressure correction and the black and green lines
marked as Tupi 1 and Tupi 3 are the muon intensities registered on  the inclined direction ($45^0$), before and after pressure correction. This figure caption is also valid for Fig.8 and Fig.10. In all cases the Tupi muon intensities are compared with the solar proton intensity in the MeV energy region, registered by the SIS detector on board of the ACE spacecraft and located in the L1 (Lagrange) point \cite{ace}. The orange and magenta lines marked as SIS 1 and SIS 2 represent the SIS integral solar proton flux for energies above 10 GeV and 30 GeV respectively.

Some important features observed in these Figures are: 

a) As already commented, the influence of the atmospheric pressure variation in the muon intensity at sea level is not relevant.\\

b) In both muon intensity levels, high and low, the vertical muon intensity is around twice higher than the muon intensity impacting with $45^0$.
This behavior is in agreement with the expected zenith dependence like $\cos^2 \theta$ for the muon intensity in the GeV region.\\

c) The transition between the low to high muon intensity and vice-verse is abrupt. In most cases, the high level muon intensity is up to 10 times higher than the low level muon intensity.\\

d) There are days where the muon intensities remain in the low level during the 24h. The opposed situation is also observed, the muon intensity remain in the high level. However, in this case, it is observed muon intensities fluctuate exactly in the ``solar window to muons'' limits
(9h and 18h local time, see Fig.8 and Fig.10). These changes, probably are associated to transient changes on the mechanism of solar energetic particle emission, modifying the solar particle spectrum, as well as due to magnetic field fluctuations in the interplanetary space that include the magnetic disturbances on the Earth's magnetic field, increasing the geomagnetic cutoff (shielding effect) or decreasing the geomagnetic cutoff (enhancement of particles with low rigidity). \\ 
 
e) In Fig.9, we have included also for comparison the solar proton flux in the keV energy region as observed by the EPAM detector on board of the ACE spacecraft and located at L1. On 2007 April 30, the Tupi telescopes have detected a muon enhancement in coincidence with the arrival of a large keV proton burst
in the EPAM detector. This means that the energy spectrum of the proton burst extend beyond 1 GeV, because they can produce muons in the earth's atmosphere. However, this proton enhancement was not registered by SIS detector (MeV region) located in the same ACE spacecraft.
In addition, the solar proton intensity in the MeV region as is observed by ACE-SIS detector has been very stable, at least during the period the observation of this survey and its small fluctuations happen in coincidence3
with the Tupi muon fluctuations as is shown in Fig.10 on 2007 April 26. 

f) The solar proton intensity as detected on ACE SIS detector (above 10-30 MeV region) is around 100 times higher than the Tupi high level muon intensity and correspond to primary particles with rigidities above 0.4 GV (0.4 GeV for protons) Under the assumption that the high intensity muons are produced by solar protons (ions), we have estimated the intensity of solar protons (ions) with rigidity above 1 GV. 
A typical representation (under quiet conditions) of the intensities (EPAM, SIS and Tupi) is presented in Fig.9 on 2007/04/27 and the integral energy spectrum is shown in Fig.11. As expected its spectrum is close to a power law, with an integral spectral index of $-0.855\pm 0.157$. This behavior of the energy spectrum strongly suggest that the Tupi high level muons are produced by solar particles from the tail of the energy spectrum.\\

g)In order to compact all observation, the hourly muon intensity at sea level for $E_{\mu}\geq 0.2\;GeV$, averaged over two months is shown in the upper panel of Fig.12, as well as,
the daily variation of the amplitude of the first harmonic for the muon intensity defined as
\begin{equation}
A=\frac{maximum\;intensity-average\;intensity}{average\;intensity},
\end{equation}
averaged over two months is shown in the lower panel.
In the ``solar window to muons'' region ($\sim 8.0h$ until  $\sim 19.0h$) there are two peaks. The first peak is linked to the amplitude and phase of the first harmonic of the diurnal anisotropy as 20\% and 13h respectively. The second peak at 18h is linked to the east-west anisotropy due to the Earth's magnetic field. The high value for the amplitude obtained here, corresponds to the minimum solar activity period and must be smaller when data obtained on the maximum solar activity period are included. 

The ``solar window to muons'' region include all the values of the maxima and minima  phase observed by neutron monitors and muon telescopes (see Table I). 
Another important  feature, and crucial for the formulation of a new picture for the origin of the solar daily variation, is the verification that in the region coverted by the ``solar window to muons'' the interplanetary magnetic lines overtake the Earth. The situation is summarized in Fig.13.  All IMF lines inside of shaded region overtake the Earth.

On the other hand, we have used the Fast Fourier Transformation (FFT) for studying periodicities and scaling properties that
might be present in time series constructed using the hourly muon intensities. The power spectra of the hourly muon intensities for two months are show in Fig.14. In this case there is a series of peaks, such as at 0.99 days (daily anisotropy) and a small peak also can be observed at 0.58 days known as the semi-diurnal anisotropy.
The harmonic $27/n$, with $n=4$ giving a peak at $\sim 7$ days and correspond to the quasi-periodic corotating streams that occurs due to solar rotation period of 27 days.  We would like point to out that the harmonics like $27/n$ has been observed in the power spectra of solar wind speed measurements reported by Burlaga and Lazarus \cite{burlaga00}. These results strongly suggest that the solar wind and the protons producing the high muon intensity have a common origin, the Sun. 

Finally, in a previous article \cite{navia05_2} results obtained with the Tupi telescope working on raster scan regime (telescope always pointed to the IMF lines) were presented, where sudden depressions in the muon intensity were observed. Initially, the events were interpreted as mini-Forbush events, a shielding effect due to the passage for the Earth vicinity of a small interplanetary disturbance.  
The data shows that the depressions in most of the cases began at $\sim 21h$ UT and that corresponds to at 18h hours (local time). Now it is clear, the depressions are the exit  of the ``solar windows for muons''. The observations correspond to 2004 and 2005 periods. Now, 2007 in the minimum of solar activity, the abrupt transitions between high and low levels of the muon intensity, are more frequent.

\section{Conclusions and remarks}

The observation of the solar daily variation of the cosmic ray intensity in a wide region of rigidities
is a challenge to phenomenological models to explain the phenomenon. Despite there are some mechanisms to explain the effect especially in the low rigidity band, none of the possibilities can explain the daily variation in all the cases. Therefore, the cosmic ray daily variation is still an open question.

The solar daily variation of the cosmic ray particle intensity is believed to be due to the corotating galactic cosmic rays with the interplanetary magnetic field lines. However, the data of solar daily variation, especially in the high rigidity region obtained by the underground secondary cosmic ray intensity  has shown remarkable changes in phase and amplitude during several decades of observation \cite{Ahluwalia97} and \cite{peacock67}, hard to explain under the assumption of the standard diffusion-convection picture of galactic cosmic rays. 

In this paper we have presented new results on the daily variations of cosmic ray muons with $E_{\mu}\geq 0.2\;GeV$ based on the data of two directional muon telescopes at sea level and with a rigidity of response to cosmic proton spectrum above 0.4 GV. Abrupt transition between a low to a high muon intensity (up to $\sim 10$ times higher) and vice-verse were observed, the period of high muon intensity is from $\sim 8.0h$ until $\sim 19.0h$ (local time) and coincides with the period in which the solar magnetic lines intercept the surface of the Earth. This behavior strongly suggest that the high muon intensity level is produced by protons whose origin is the Sun and implying that the solar proton energy spectrum extends beyond 1 GeV. 

This assumption is reinforced by some important experimental features such as
the solar proton energy spectrum obtained by direct observations by detectors on board of the ACE spacecraft, EPAM protons(keV band) and SIS protons (MeV band) and which can be extended to higher energies following a power law by the Tupi protons (GeV band) on the basis of the high muon intensity. 
The FFT power spectra of the hourly muon intensities have a series of peaks, such as the harmonic $27/n$, with $n=4$ giving a peak at $\sim 7$ days and correspond to the quasi-periodic corotating streams that occurs due to solar rotation period of 27 days. These same harmonics have been found in the power spectra of solar wind speed measurements, suggesting a common origin by the solar wind and the protons producing the high muon intensity. 

The assumption of a solar origin for energetic particles, as responsible for the diurnal intensity variation, takes into account also the phase shift of diurnal variation at reversal of the solar magnetic field. If the origin of the energetic charged particle is the Sun, their propagation follows the interplanetary magnetic field lines. In addition, the structure of the interplanetary magnetic field depends on the expansion speed of the plasma (solar wind) expelled by the Sun. Fig.13 summarized the situation, where the solar magnetic field lines between the Sun and the Earth for several values of the speed of solar wind is shown \cite{helios}.   
In the period of low solar activity,  the average  speed of the solar wind increases, and the magnetic lines and consequently the energetic charged solar particle propagation is close to the Earth-Sun line. The result is observed as a shift of the phase of the anisotropy to early hours \cite{Ahluwalia97}.   

The so called ``solar window to muons'' ($\sim 8h$ until $\sim 19h$) includes all values of the anisotropy phase, observed in a wide range of rigidity (see Table I). However, some results \cite{jacklyn63} indicate that the maximum amplitude in the range of (1-100)GV, observed with narrow-angle detectors at the equator, would be 0.4 per cent of the average primary flux and independent of the rigidity. This disagrees with our results where an amplitude of $\sim 20\%$ is observed (see Fig.11). Even so, it must be smaller when results obtained in the high solar activity period are included. The high amplitude observed in this work can be a consequence of the brazilian magnetic anomaly region, where the lowest magnetic region located at 26S, 53W, is very close to the position 22S, 43W where the Tupi experiment is located. The contours of the omnidirectional proton intensity ($E_p>10$ MeV) around the anomaly region is ploted in Fig.14 \cite{nichitiu}. The lowest magnetic field in the region of the anomaly is three times lower than the magnetic field at polar regions.    

Our data also suggested that the amplitude of the diurnal variation would be dependent on the rigidity. This conclusion is in accordance with previous results where the amplitudes decrease systematically with increasing rigidity \cite{Ahluwalia97}. The second peak in the amplitude at $\sim 18h$ (see Fig.12 lower panel) is linked with the east-west anisotropy due to Earth's magnetic field, because particles of rigidity below 60 GV are subject to the geomagnetic effect.

We would like to point out, that the Tupi results are on the basis of two muon telescopes with a high sensitivity, capable to observer, under certain conditions, muon enhancement in association with solar flares of small scale (type C) and muon enhancements in coincidence with direct observation of solar protons EPAM and SIS on board of the ACE spacecraft. 

\section{Acknowledments}

This work is supported by the National Council for Research (CNPq) in Brazil, under Grant No. $479813/2004-3$. 
The authors wish to express their thanks to Dr.
A. Ohsawa from Tokyo University for help in the first stage of the
experiment. We are also grateful to the various catalogs available  on the web and to 
their open data police, especially to the ACE Real-Time Solar Wind (RTSW) Data.

\newpage

\begin{table}
\vspace*{+3.0cm} 
\caption{The times for the maxima and minima of the diurnal variation (phase) observed in four global sites.
For the 1965 to 1994 period (Deep River, Huancayo and Embudo) and for the 1971 to 1994 period (Nagoya)}
\begin{ruledtabular}
\begin{tabular}{ccccc}
 Detector & Type & Mean Rigidity (GV) &  Minimum Phase (hour-LT)& Maximum Phase (hour-LT)\\
\hline

Deep River\footnotemark[1] & Neutron Monitor & 16 & 12.5 & 16.0\\
Huancayo  \footnotemark[1] & Neutron Monitor & 33 & 7.6 & 14.4\\
Embudo    \footnotemark[1] & Muon Telescope  & 134 & 8.3 & 17.7\\
Nagoya    \footnotemark[2] & Muon Telescope  &  60 & 11.0 & 12.8\\
\end{tabular}
\end{ruledtabular}
\footnotetext[1]{Ref.\cite{Ahluwalia97}.}
\footnotetext[2]{Ref.\cite{mori96}.}
\end{table}

\vspace*{+10.0cm}

\newpage
\newpage
\vspace*{+8.0cm} 

\begin{figure}[th]
\vspace*{-9.0cm}
\includegraphics[clip,width=0.7
\textwidth,height=0.7\textheight,angle=0.] {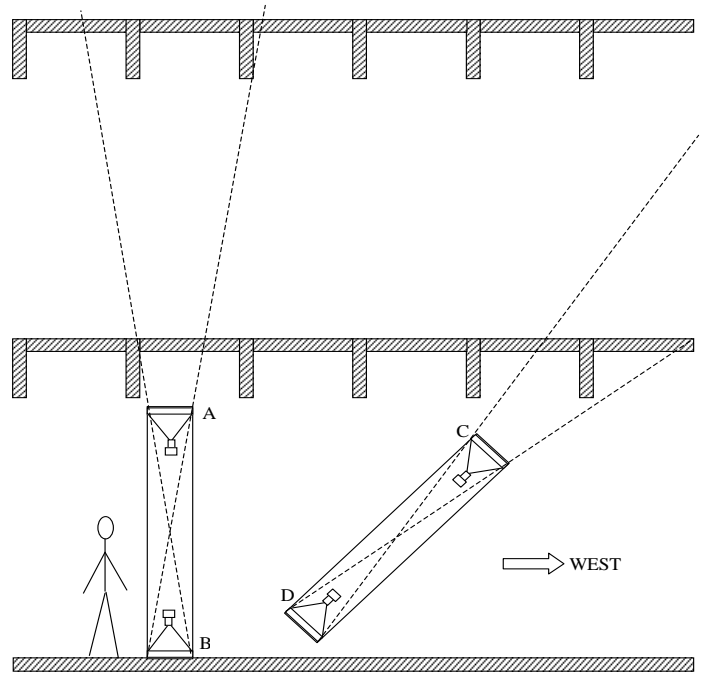}
\vspace*{-5.0cm}
\caption{Experimental setup of the Tupi experiment Phase II, showing the two telescopes.}
\end{figure}

\begin{figure}[th]
\vspace*{-3.0cm}
\includegraphics[clip,width=0.9
\textwidth,height=0.9\textheight,angle=0.] {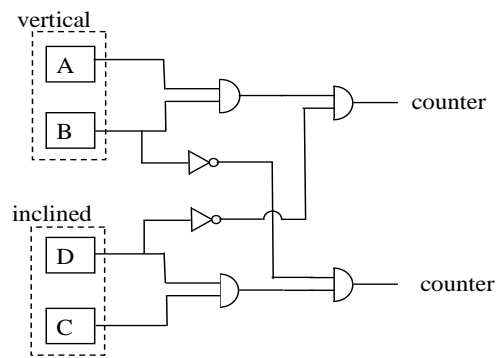}
\vspace*{-13.0cm}
\caption{General layout of the logic implemented in the data acquisition system. The vertical telescope uses 
 a veto or anti-coincidence guard system with a detector of the inclined telescope and vice-verse. This system allow only the detection of muons traveling close to the telescope axis direction.}
\end{figure}

\vspace*{+8.0cm} 

\begin{figure}[th]
\vspace*{-8.0cm}
\includegraphics[clip,width=0.4
\textwidth,height=0.4\textheight,angle=0.] {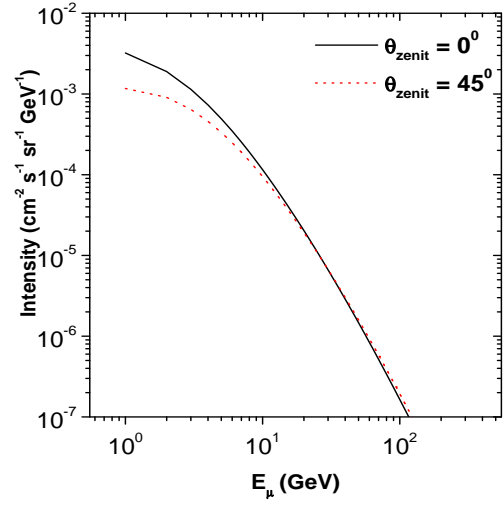}
\vspace*{-2.0cm}
\caption{Expected differential energy spectrum of muons at sea level for two zenith angles. }
\end{figure}

\begin{figure}[th]
\vspace*{-1.0cm}
\includegraphics[clip,width=0.7
\textwidth,height=0.7\textheight,angle=0.] {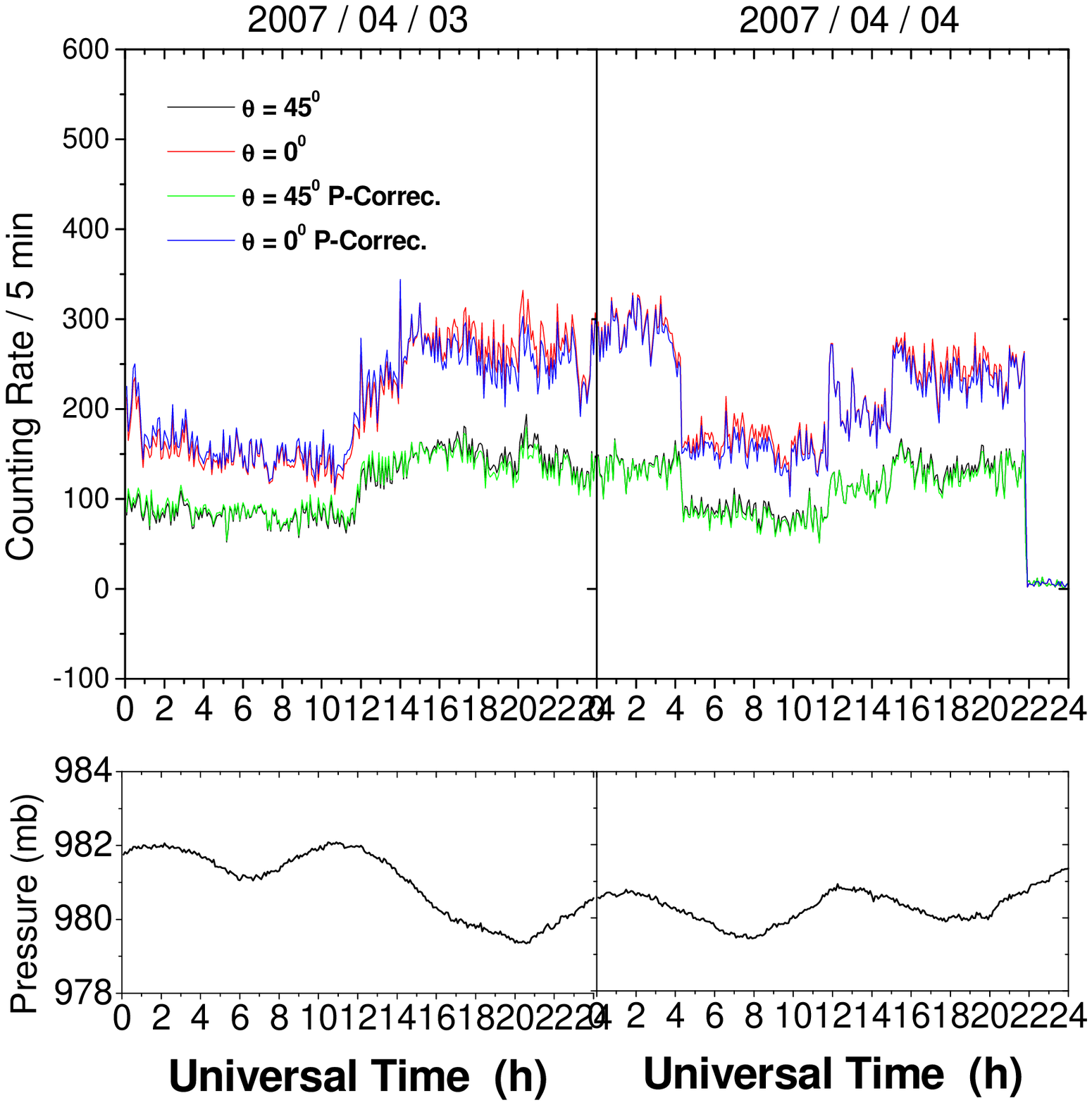}
\vspace*{-4.0cm}
\caption{Upper panel: The X-ray flux according GOES11 for two wave length band. Lower panel:
The 5 minutes Tupi integral muon intensity, the blue and red lines are for the vertical direction and
green and black lines are for inclines direction. The red and black represent pressure corrected (PC) rate.}
\end{figure}

\begin{figure}[th]
\vspace*{-1.0cm}
\includegraphics[clip,width=0.5
\textwidth,height=0.5\textheight,angle=0.] {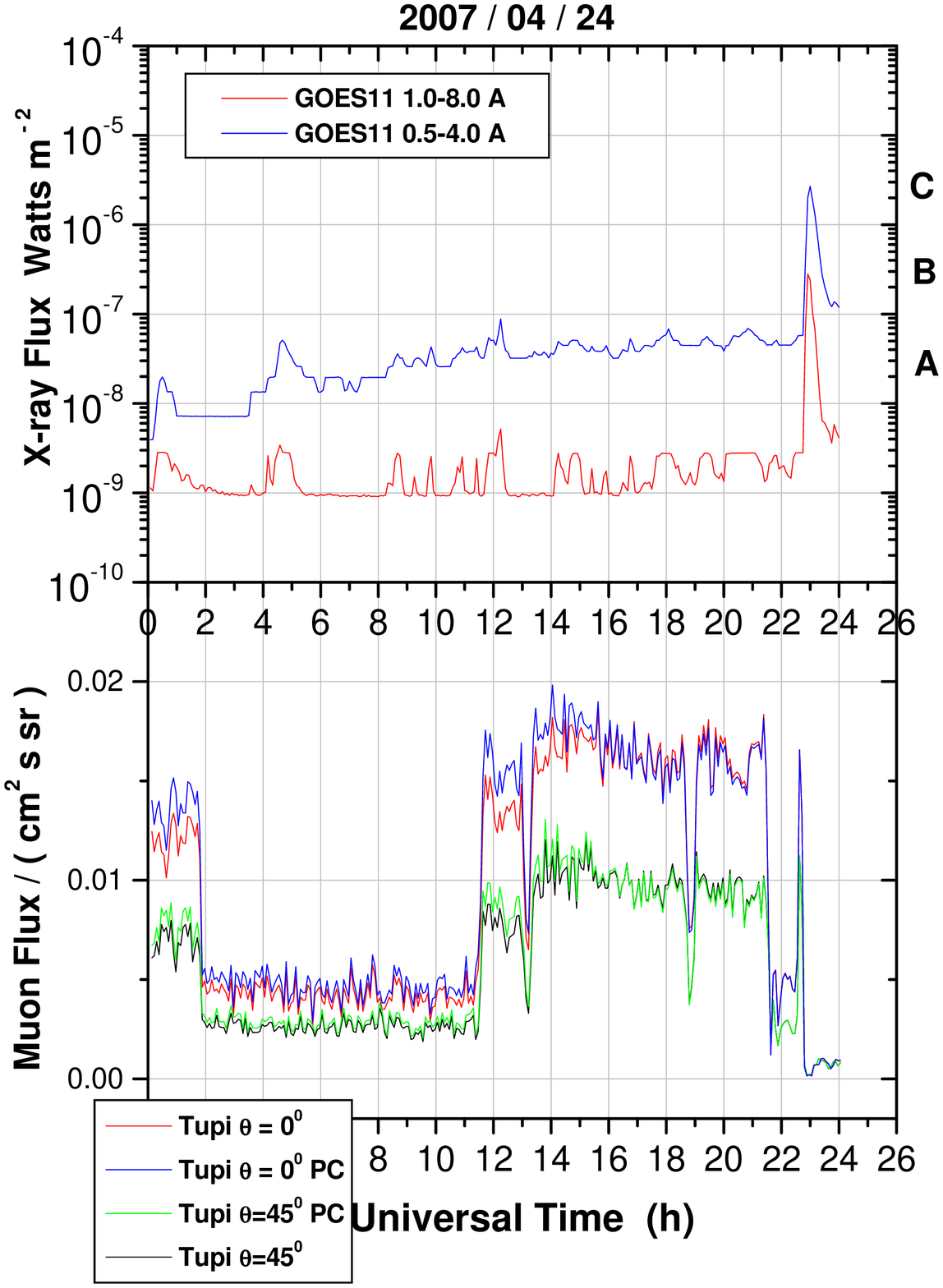}
\vspace*{-0.5cm}
\caption{Upper panel: The X-ray flux according GOES11 for two wave length band. Lower panel:
The 5 minutes Tupi integral muon intensity, the blue and red lines are for the vertical direction and
green and black lines are for inclines direction. The red and black represent pressure corrected (PC) rate.}
\end{figure}

\begin{figure}[th]
\vspace*{-7.8cm}
\hspace*{-1.0cm}
\includegraphics[clip,width=1
\textwidth,height=1\textheight,angle=0.] {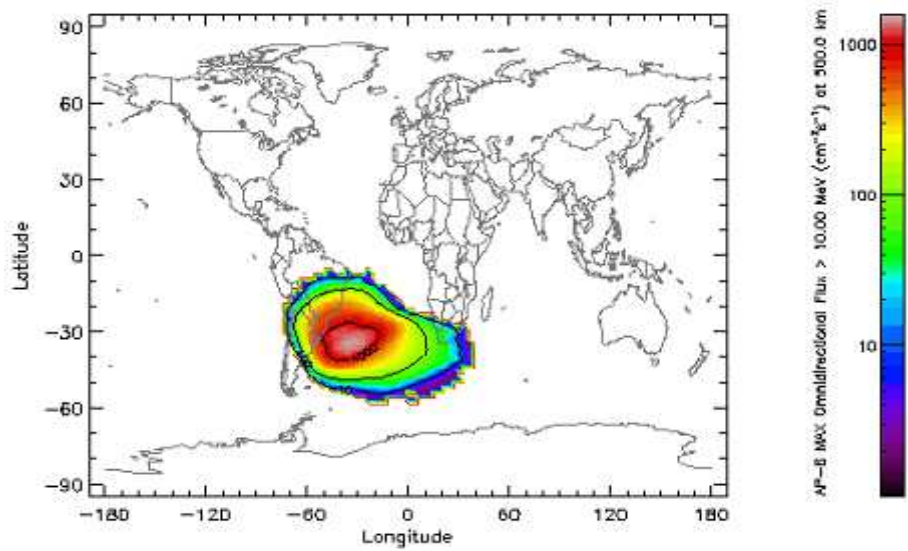}
\vspace*{-8.5cm}
\caption{The omni directional proton intensity $E>10$ MeV, around the brazilian magnetic anomaly. Figure from ref. \cite{nichitiu}}
\end{figure}

\newpage

\vspace*{+8.0cm} 

\begin{figure}[th]
\vspace*{-10.0cm}
\includegraphics[clip,width=0.9
\textwidth,height=0.9\textheight,angle=0.] {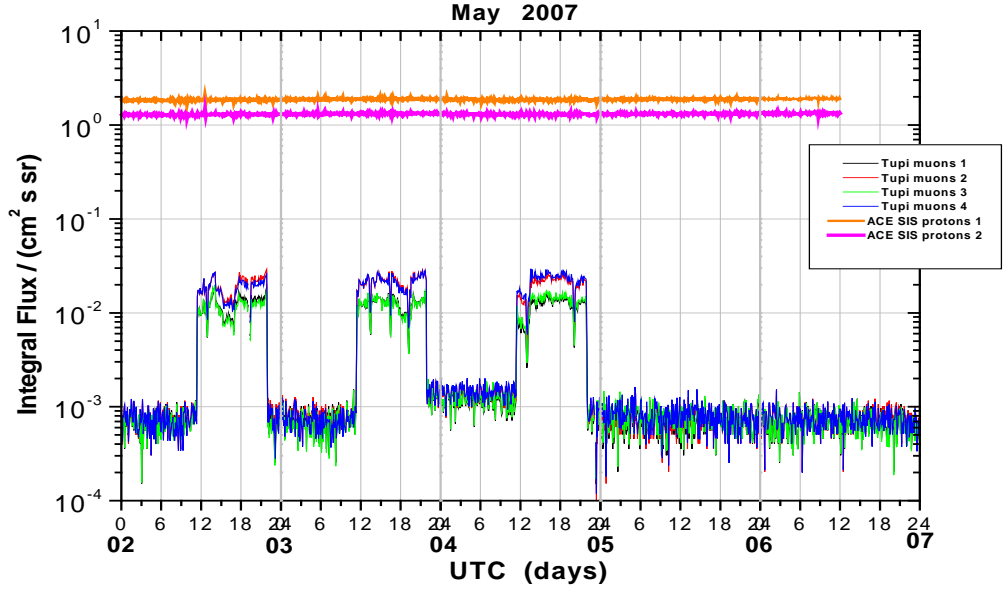}
\vspace*{-11.0cm}
\caption{The 5 minutes integral flux: The orange and magenta marked as SIS 1 and SIS 2 represent the SIS integral solar proton flux for energies above 10 GeV and 30 GeV respectively.
The blue and red lines marked as Tupi 4 and Tupi 2 respectively are the muon intensity registered in the vertical direction, before and after atmospheric pressure correction and the black and green lines
marked as Tupi 1 and Tupi 3 are the muon intensity registered on  the inclined direction ($45^0$), before and after pressure correction. In all cases the muon energy threshold is $E_{\mu}\geq 0.2\;GeV$.}
\end{figure}

\begin{figure}[th]
\vspace*{-2.0cm}
\includegraphics[clip,width=0.9
\textwidth,height=0.9\textheight,angle=0.] {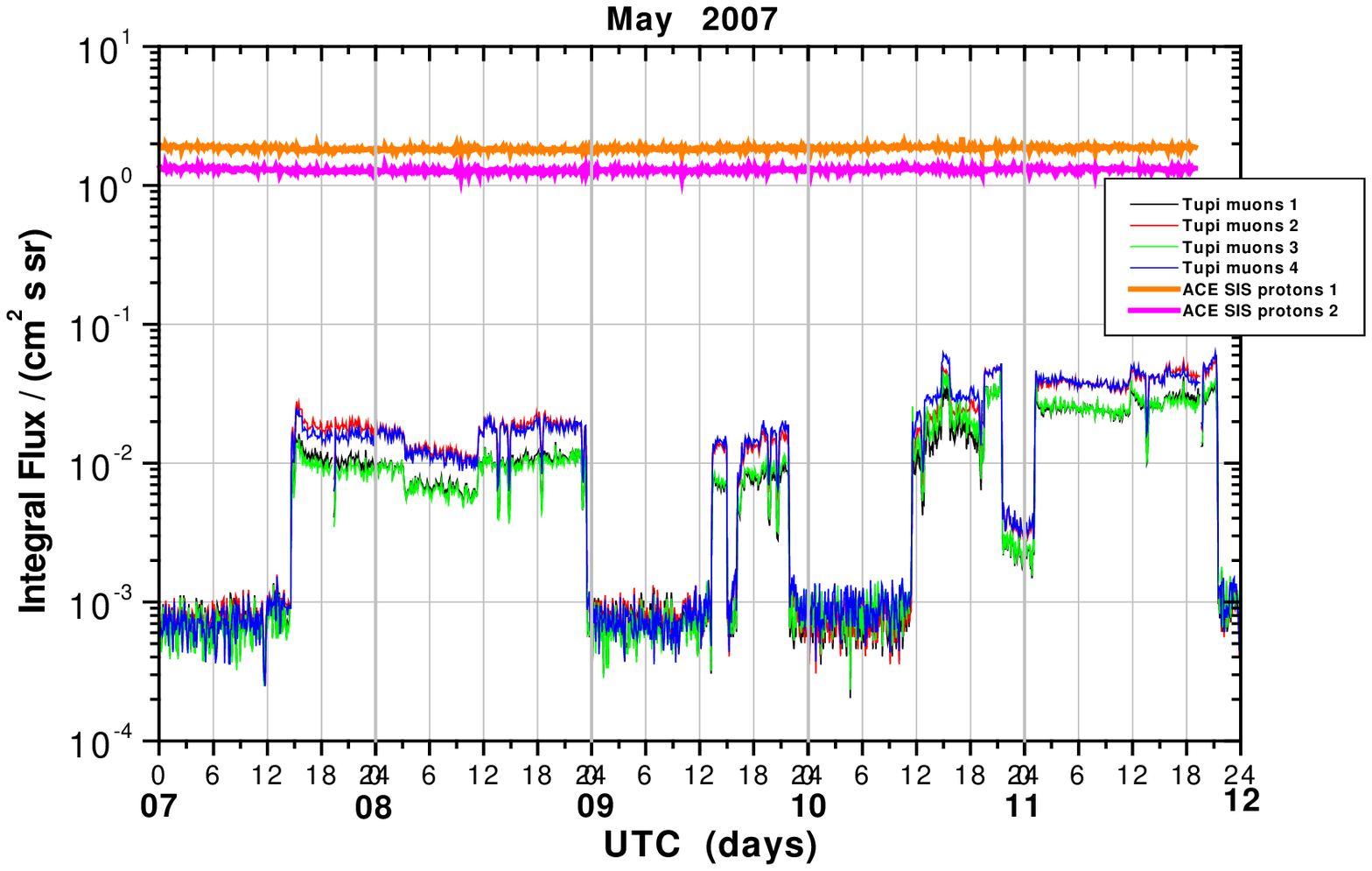}
\vspace*{-11.0cm}
\caption{The same as figure Fig.5.}
\end{figure}

\newpage

\vspace*{+8.0cm} 

\begin{figure}[th]
\vspace*{-10.0cm}
\includegraphics[clip,width=0.9
\textwidth,height=0.9\textheight,angle=0.] {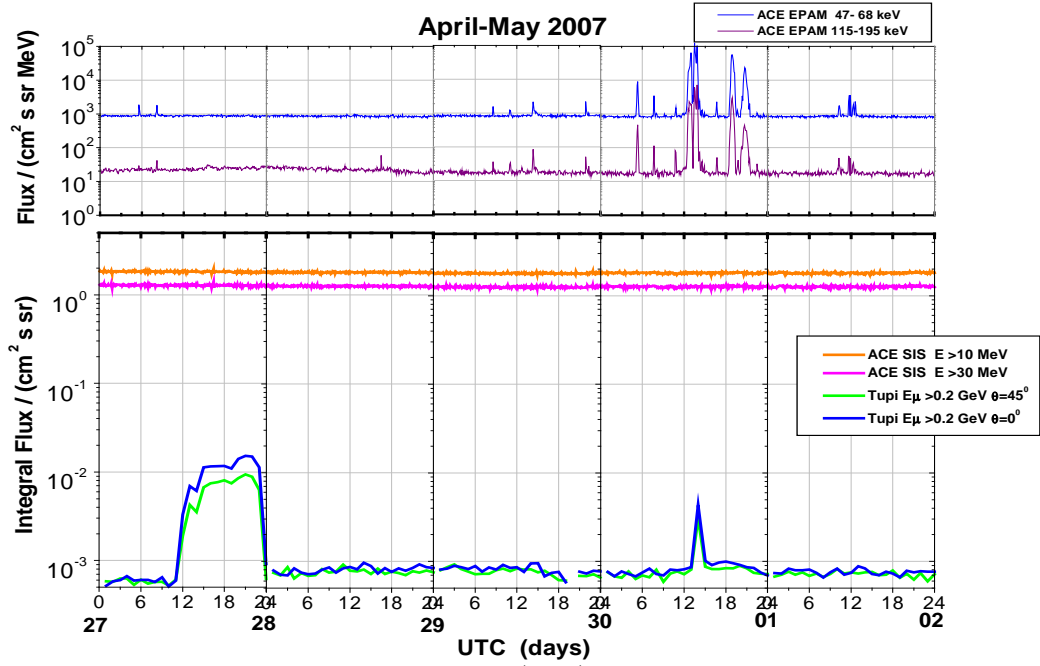}
\vspace*{-11.0cm}
\caption{Upper panel: The 5 minutes differential flux of EPAM (ACE) protons, for two keV energy  bands. 
Lower panel:The orange and magenta represent the 5 minutes list SIS integral solar proton flux for energies above 10 GeV and 30 GeV respectively.
The blue and green lines are the hourly muon integral intensity registered in the vertical and inclined directions
respectively, after atmospheric pressure correction.  In all cases the muon energy threshold is $E_{\mu}\geq 0.2\;GeV$. }
\end{figure}

\begin{figure}[th]
\vspace*{-0.0cm}
\includegraphics[clip,width=0.9
\textwidth,height=0.9\textheight,angle=0.] {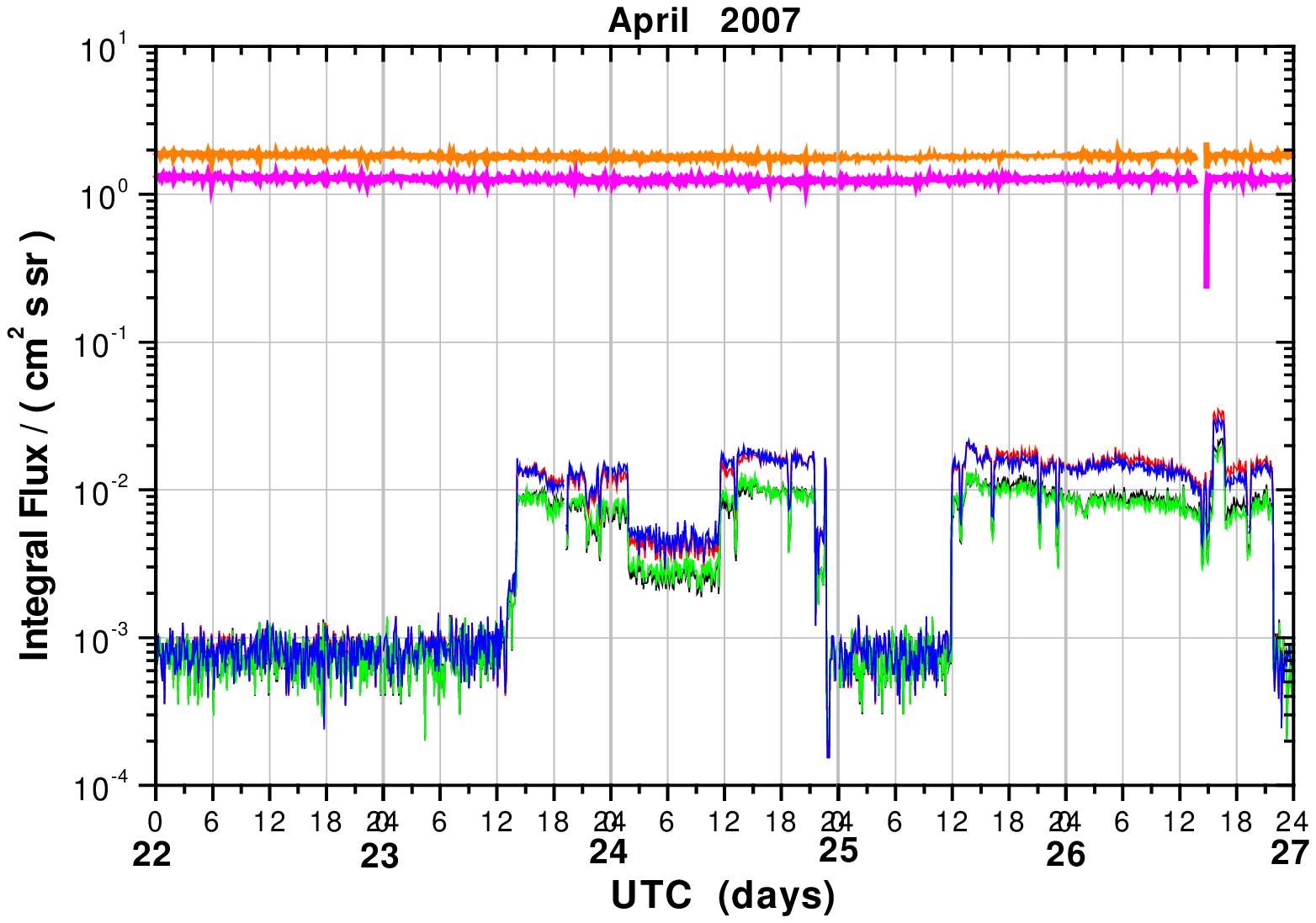}
\vspace*{-11.0cm}
\caption{The same as figure Fig.5 }
\end{figure}

\newpage

\vspace*{+8.0cm}

\begin{figure}[th]
\vspace*{-10.0cm}
\includegraphics[clip,width=0.5
\textwidth,height=0.5\textheight,angle=0.] {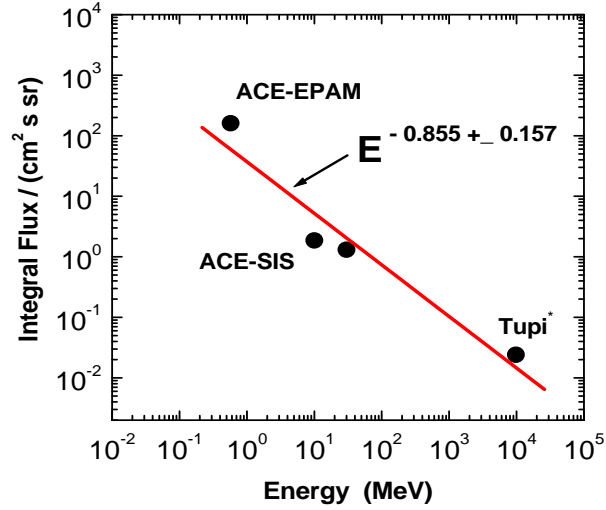}
\vspace*{-4.0cm}
\caption{The integral solar proton flux, under the assumption of that the Tupi high level muon intensity is produced by solar protons}
\end{figure}

\begin{figure}[th]
\vspace*{-2.0cm}
\includegraphics[clip,width=0.7
\textwidth,height=0.7\textheight,angle=0.] {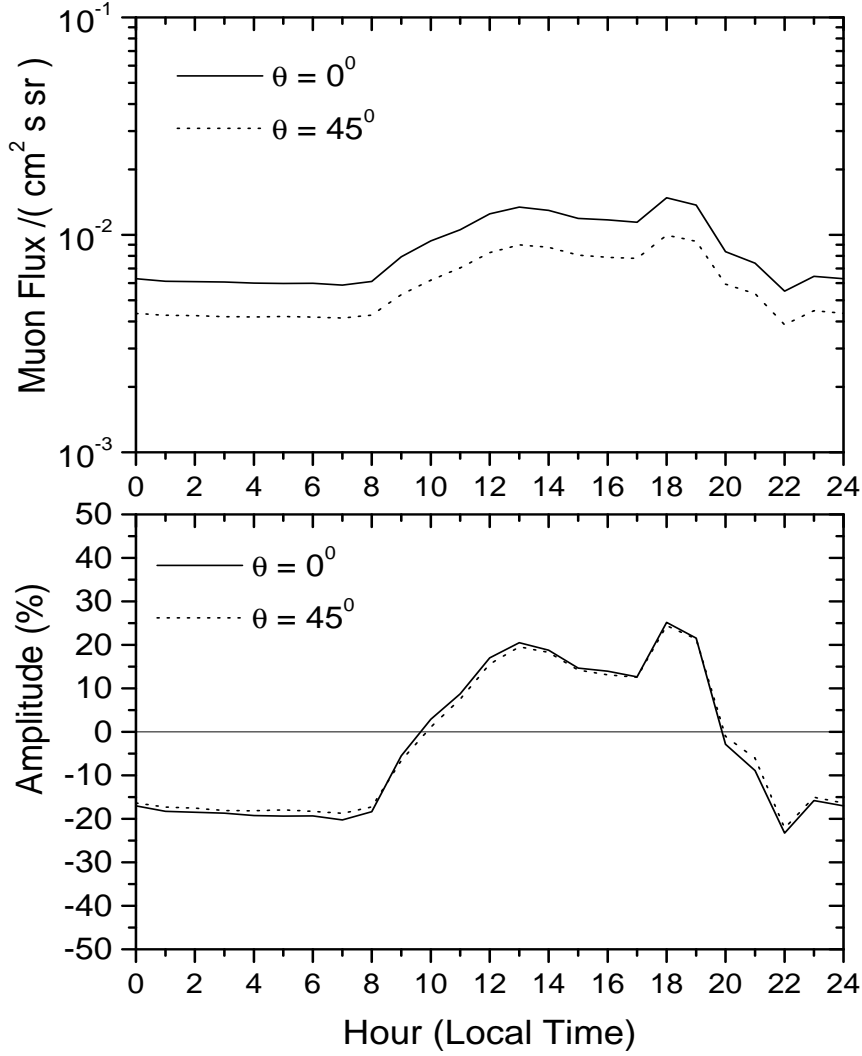}
\vspace*{-1.0cm}
\caption{Upper panel: The hourly muon intensity at sea level for $E_{\mu}\geq 0.2\;GeV$, averaged over two months.
Lower panel: The amplitude of the first harmonic for the muon intensity,  averaged over two months.}
\end{figure}

\newpage

\begin{figure}[th]
\vspace*{-3cm}
\includegraphics[clip,width=0.85
\textwidth,height=0.85\textheight,angle=0.] {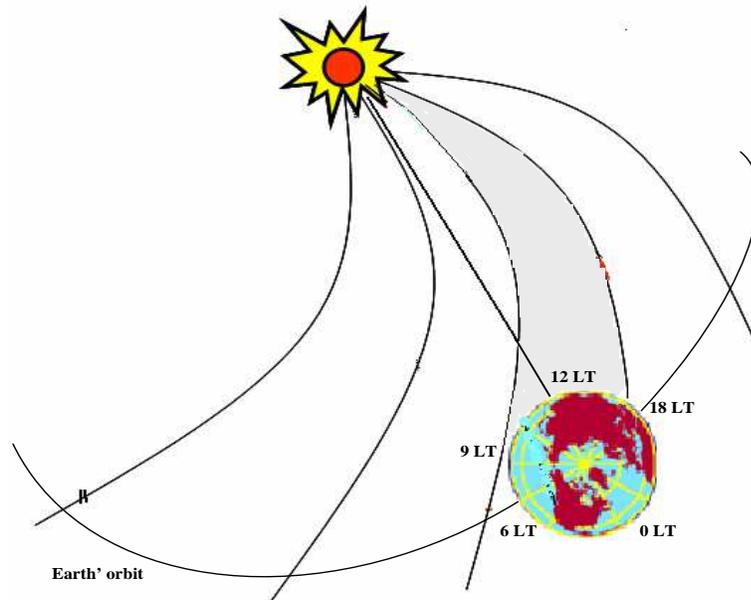}
\vspace*{-10.0cm}
\caption{The solar field lines in the ecliptic plane. The field lines inside the shared region overtake the Earth.}
\end{figure}

\begin{figure}[th]
\vspace*{-1.0cm}
\includegraphics[clip,width=0.5
\textwidth,height=0.5\textheight,angle=0.] {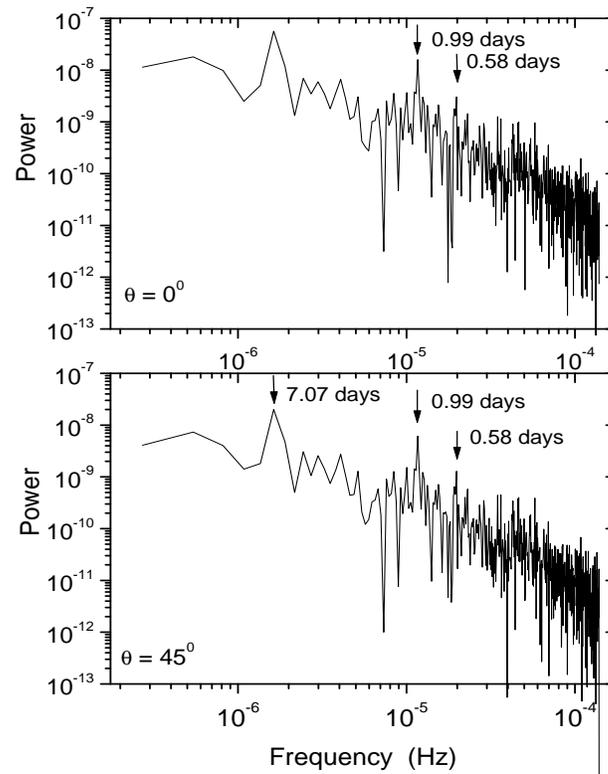}
\vspace*{-1.0cm}
\caption{The power spectral density as a function of the frequency for the hour averages of muon intensity measured during two moths.}
\end{figure}

\newpage

\begin{figure}[th]
\vspace*{-4.0cm}
\hspace*{-5.0cm}
\includegraphics[clip,width=1.3
\textwidth,height=1.3\textheight,angle=0.] {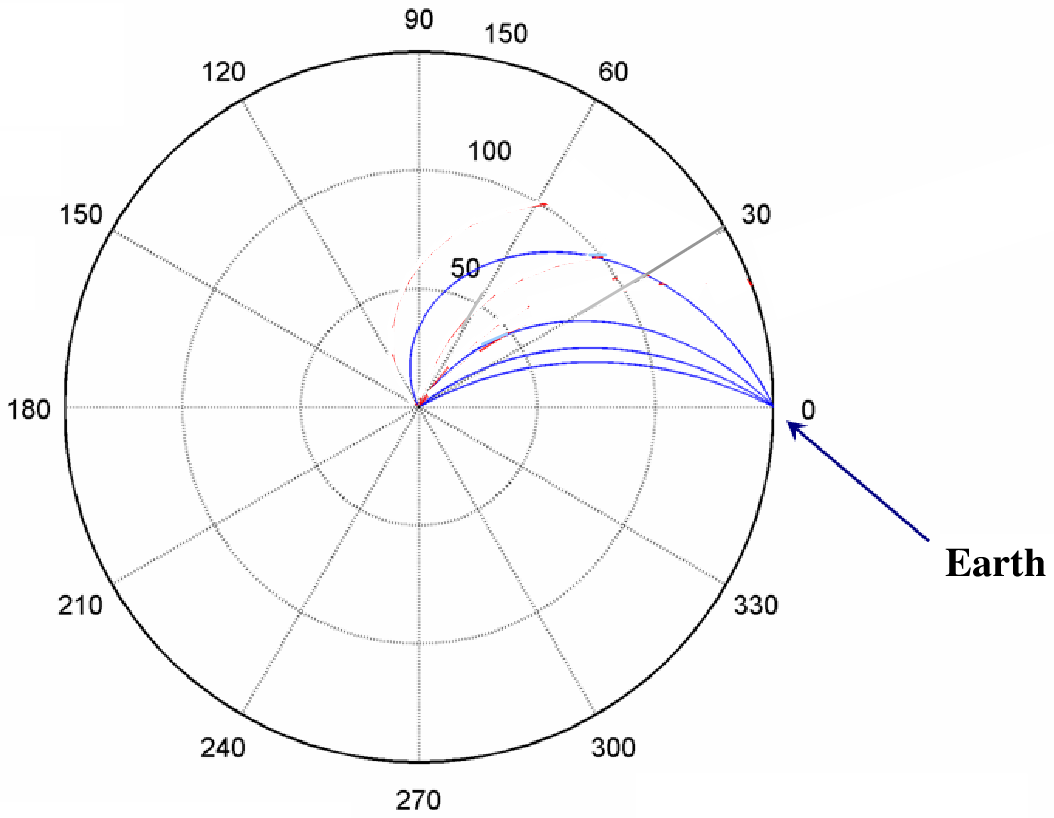}
\vspace*{-18.0cm}
\caption{The structure of the interplanetary magnetic field as a function of the 
expansion speed of the solar wind plasma expelled by the Sun, with solar wind speed (from top to bottom): 200, 400, 600 and 800 km/s. Figure from ref. \cite{helios}}
\end{figure}


\begin{references}
\bibitem{parker63}E. N. Parker, Interplanetary Dynamical Processes, Monographs and Texts
in Phys. and Astro., 8, Inter-science  Publishers: New York, New York (1963). 
\bibitem{simpson54}J. Simpson, Phys. Rev. 94, 426 (1954).
\bibitem{richardson04} I. A. Richardson, Space Science Rev., 111, 267 (2004). 
\bibitem{forbush37}S. C. Forbush, Phys. Rev. 51, 1108 (1937).
\bibitem{iucci79}N. Iucci, M. Parise, M. Storine and G. Villoresi, Il Nuovo Cimento, 2C, 1 (1979). 
\bibitem{forman75} A. Forman, L. J. Gleeson, Astrophys. Space Sci. 32, 77. (1975).
\bibitem{kota97}J. Kota, Proc. 25th Int. Cosmic Ray Conf.  2, 145 (1997).
\bibitem{augusto03}C. R. Augusto, C. E. Navia and M. Robba, Nucl. Instrum. Methods Phys. Res. 503, 554 (2003).
\bibitem{navia05}C. E. Navia, C. R. A. Augusto, M. B. Robba, M. Malheiro and H. Shigueoka, Astrophys. Journal. 621, 1137 (2005).
\bibitem{augusto05} C. R. A. Augusto, C. E. Navia and M. B. Robba, Phys. Rev. D71, 103011 (2005).
\bibitem{green79}P. J. Green et al., Phys. Rev. D20, 1598 (1979).
\bibitem{lipari93}P. Lipari, Astr. Par., 1, 195 (1993). 
\bibitem{nichitiu} F. Nichitiu, available in\\
$http://www.Inf.infn.it/seminars/nichitiu.ppt$
\bibitem{Ahluwalia97} H. S. Ahluwalia and M. M. Fikani, Proc. 25th Int. Cosmic Ray Conf.  2, 125 (1997).
\bibitem{mori96}S. Mori, Il Nuovo Cimento, Vol.19, 791 (1996).
\bibitem{ace}ACE Real-Time Solar Wind Data Website, World Wide Wide: http://www.sec.noaa.gov/ace/
\bibitem{burlaga00}L. F. Burlaga and A. J. Lazarus, J. Geophys. Research, V.105, 2357 (2000).
\bibitem{navia05_2}C. E. Navia, C. R. A. Augusto, K. H. Tsui and M. B. Robba, Phys. Rev. D72, 103001 (2005).
\bibitem{peacock67} D. S. Peacock and  T. Thanbyahpillai, Nature 215, 146 (1967).
\bibitem{helios}Helios Vocca Group, Solar Magnetic field through Lisa and Earth, available in\\
$http://www.fisica.unipg.it/vocca/CCLRC.pdf$.
\bibitem{jacklyn63}R. M Jacklyn, Nature 200, 1306 (1963).


\end{references}
\end{document}